\renewcommand{\Re}{\mathop{\rm Re}\nolimits}
\renewcommand{\Im}{\mathop{\rm Im}\nolimits}
\documentstyle[12pt,amssymb]{article}
\begin{document}
\begin{titlepage}
\begin{flushright}
UM-TH-95-05\\
March 1995\\
\end{flushright}
\vskip 2cm
\begin{center}
{\large\bf Eliminating the Hadronic Uncertainty}
\vskip 1cm
{\large Robin G. Stuart}
\vskip 1cm
{\it Randall Physics Laboratory,\\
 University of Michigan,\\
 Ann Arbor, MI 48190-1120,\\
 USA\\}
\bigskip
and\\
\bigskip
{\it Instituto de F\'\i sica,\\
 Universidad Nacional Aut\'onoma de M\'exico,\\
 Apartado Postal 20-364, 01000 M\'exico D. F.\\
 MEXICO}\\
\end{center}
\vskip .5cm
\begin{abstract}
The Standard Model Lagrangian requires the values of the fermion masses,
the Higgs mass and three other experimentally
well-measured quantities as input in order to become predictive. These
are typically taken to be $\alpha$, $G_\mu$ and $M_Z$. Using the first
of these, however introduces a hadronic contribution that leads to
a significant error. If a quantity could be found that was measured at
high energy with sufficient precision then it could be used to replace
$\alpha$ as input. The level of precision required for this to happen
is given for a number of precisely-measured observables. The $W$ boson
mass must be measured with an error of $\pm13$\,MeV,
$\Gamma_Z$ to $0.7$\,MeV and polarization asymmetry, $A_{LR}$, to
$\pm0.002$ that would seem to be the most promising candidate.
The r\^ole of renormalized parameters in perturbative calculations is
reviewed and the value for the electromagnetic coupling constant
in the $\overline{\rm MS}$ renormalization scheme that is consistent with
all experimental data is obtained to be
$\alpha^{-1}_{\overline{\rm MS}}(M^2_Z)=128.17$.

\end{abstract}
\end{titlepage}

\setcounter{footnote}{0}
\setcounter{page}{2}
\setcounter{section}{0}
\newpage

A renormalizable theory generally contains a number of {\it a priori\/}
free parameters whose numerical values must be fixed by independent
input; one for each parameter. Then and only then does the model
become predictive. In the case of the Standard Model of electroweak
interactions, the free parameters include the fermion masses, $m_f$, gauge
couplings, weak boson ($W$ and $Z$) masses and the Higgs mass. In
fact this list constitutes an over-complete set and any one of these
may be predicted knowing the values of the rest.

In order to make the most precise predictions from the model, one chooses
the best determined experimental parameters as input. For most purposes
the fermion masses are sufficiently well known as to not present any
limitation in precision. In addition, they usually enter as terms of
${\cal O}(\alpha m_f^2/q^2)$ and are thus usually negligible for
$q^2\gtrsim M_W^2$. The Higgs mass is, of course, undetermined
and allowed to float in calculations. The top quark is mass poorly known
\cite{CDF,D0}
but for processes that do not contain top quarks in the final state
its dependence is suppressed by $\alpha$. The remaining three pieces
of experimental input required to make the model predictive are usually
taken to be those experimentally extremely well-measured quantities,
\begin{eqnarray}
\alpha^{-1}&=&137.0359895\pm0.0000061\label{eq:firstexp}\\
G_\mu&=&(1.16639\pm0.0002)\times 10^{-5}\,{\rm GeV}^{-2}\\
M_Z&=&91.187\pm0.007\,{\rm GeV}\label{eq:lastexp}
\end{eqnarray}
Note that in principle all these parameters can be extracted from
experiments without detailed knowledge of the underlying model. For this
reason we will refer to them as model-independent physical observables.
The electromagnetic coupling constant, $\alpha$, can be extracted from
the cross-section for Thomson scattering by applying the
{\sl definition\/}
\begin{equation}
\sigma_T=\frac{8\pi}{3}\frac{\alpha^2}{m_e^2}
\end{equation}
where $m_e$ is the electron mass.

The muon decay constant may be obtained from the muon lifetime,
$\tau_\mu$, by the
relation, again a definition,
\begin{equation}
\tau_\mu^{-1}=\frac{G_\mu^2 m_\mu^5}{192\pi^3}
              \left(1-\frac{8m_e^2}{m_\mu^2}\right)
              \left\{1+\frac{3}{5}\frac{m_\mu^2}{M_W^2}+
                     \frac{\alpha}{2\pi}\left(\frac{25}{4}-\pi^2\right)
                     \left(1+\frac{2\alpha}{3\pi}\ln\frac{m_\mu}{m_e}\right)
              \right\}
\end{equation}
in which $m_\mu$ and $M_W$ are the muon and $W$ boson masses.

The definitions, although motivated by our knowledge of the underlying
dynamical theory, could be applied to extract $\alpha$ and $G_\mu$
even without it or in the case that the model describing these processes
should alter radically. The mass of an unstable particle like the $Z^0$
is more problematic to define \cite{Willenbrock,Stuart1}.
Assuming only the analyticity of the
$S$-matrix, it can be shown that the model-independent physical
observable associated with the particle is the position of the complex
pole of the $S$-matrix element, $s_p$, for processes in which the
unstable particle appears in an intermediate state such as
$e^+e^-\rightarrow Z^0 \rightarrow f\bar f$. In expressions for
physically observable processes, however, the real and imaginary
parts of $s_p$ always appear together and hence any separation into
real and imaginary parts is artificial. The mass of an unstable
particle may be equally well defined as $\Re\sqrt{s_p}$ or
$\sqrt{\Re s_p}$. Both are arbitrary but both are simply related
to the model-independent physical observable, $s_p$. The pole
position, $s_p$, is a model-independent physical observable because
it may be extracted from the measurement of the cross-section
$\sigma(e^+e^-\rightarrow f\bar f)$ by means of dispersion relations
and analytic continuation. The $Z^0$ boson mass extracted and
published by the LEP experiments is based on the renormalized
mass in the on-shell renormalization scheme. This
is gauge-dependent as it is fully entitled to be. In order to obtain
close numerical agreement with the published values, Sirlin
\cite{Sirlin1} redefined the $Z^0$ mass to be
\[
M_Z=\sqrt{\Re s_p+\frac{(\Im s_p)^2}{\Re s_p}}.
\]
This, however, can no longer play the r\^ole of a self-consistent
renormalized mass. In order to make contact with experimental
data we will employ this definition in the analysis despite its
unnaturalness.

One must take care to distinguish these physical quantities from
renormalized parameters. The latter are calculational bookkeeping
devices that arise from the manner in which perturbative expansions
are performed. These quantities are decidedly unphysical and may
depend on arbitrary parameters such as the scale $\mu$ in the
$\overline{\rm MS}$ renormalization scheme or may be gauge-dependent
as they are in the on-shell renormalization scheme. This feature of the
renormalized parameters is not really a disadvantage although it
has been represented as such \cite{Sirlin1}. Any dependence on
unphysical parameters must cancel out in physical matrix elements
and can serve as a useful check.

Of the experimental input data, $\alpha$, is the best-determined.
This accuracy is misleading for calculating processes occurring at
high energies. The input value of $\alpha$ appears in association
with photon vacuum corrections that are often interpreted as the
running of $\alpha$. In this running from zero energy the hadronic
resonance region is crossed which has the effect of introducing
non-perturbative contributions. Note that a similar thing occurs
for $G_\mu$ but the conventional wisdom, based on perturbation
theory arguments, is that these contributions are suppressed by
factors $m_f^2/M_W^2$, where $m_f$ is a light fermion mass.

For $\alpha$ the non-perturbative effects are obtained via
dispersion relations from the experimentally measured cross-section
$\sigma(e^+e^-\rightarrow{\rm hadrons})$. Experimental error ultimately
degrades the effective accuracy of $\alpha$ to 8 parts in $10^4$, making
it the least accurate of the three pieces of experimental input.
The prospects for significant improvements in the experimentally
measured cross-section are rather limited, at least in the near
future.

It is a property of a renormalizable theory that
any set of experimental data containing the requisite number of
measurements of independent physical quantities can be used to make
the model predictive. In this letter we examine the sensitivity of
various precisely determined observables to the hadronic uncertainty
and thereby determine a what level of accuracy they can serve to
replace $\alpha$ as experimental input. It is to be expected that
by using experimental input that has all been measured well above the
hadronic resonance region, the need for running parameters across
the resonance region will be eliminated along with the influence of
non-perturbative effects. Of course, using a measured quantity in this
way has the apparent disadvantage that one looses an independent
observable that could have been used to test the model. In fact
what actually happens is that $\alpha$ moves from being an input
quantity to being a predicted one to be included in simultaneous fits
to data to be appropriately weighted according to its error.

The standard set of experimental input $\alpha$,
$G_\mu$ and $M_Z$ have the property that they are simply related
to very precisely measured to model-independent physical observables.
Also, in first approximation they are quite closely related to the
couplings and masses that appear in the Standard Model lagrangian;
\[
\alpha=\frac{g^2\sin^2\theta_W}{4\pi},
\hbox to 2cm{}
G_\mu=\frac{g^2}{2\sqrt{2}M_W^2}.
\]
This makes them intuitively appealing for use as experimental input
but is not at all an essential feature. For example, the peak
cross-section for $e^+e^-\rightarrow\mu^+\mu^-$ could be equally
well used as input but its lowest order approximation is
\[
\sigma=\frac{12}{\pi}\frac{\Gamma_e\Gamma_\mu}{M_Z^2\Gamma_Z^2}
\]
that is expressed in terms of partial and total widths. These may be
regarded as derived quantities since they do not themselves appear in
the lagrangian.

The ideal scenario would be to find a precise experiment that yielded
$\alpha(M_Z^2)$ directly in the way that Thomson scattering yields
$\alpha(0)$. The problem is that no such experiment exists. Any
experiment performed in the $Z^0$ resonance region will yield a
result that depends on masses and coupling constants all intermixed
in a complicated way. Thus whereas $\alpha(0)$ is justifiably
regarded as a model-independent physical observable because of its
direct association with Thomson scattering, $\alpha(M_Z^2)$ that
cannot be isolated in this way, is merely a convenient theoretical
construct without physical meaning. Analyses proporting to
give a numerical value for the latter quantity should therefore
be treated with considerable care.

For our purposes here we will work to one-loop accuracy. General
expressions in any renormalization scheme
for $\alpha$, $G_\mu$ and $M_Z$ have already been given in
ref.\cite{Stuart2,Z0POLE}.

\begin{eqnarray}
\sqrt{4\pi\alpha}&=&e\left(
         1+\frac{1}{2}\Pi_{\gamma\gamma}^{(1)\prime}(0)
          +\frac{s_\theta}{c_\theta}\frac{\Pi_{Z\gamma}^{(1)}(0)}{M_Z^2}
               +s_\theta^2\frac{\delta g}{g}
               +c_\theta^2\frac{\delta g^\prime}{g^\prime}\right)
\label{eq:firstrc}\\
\frac{G_\mu}{\sqrt{2}}&=&\frac{g^2}{8M_W^2}(1+\Delta r)
\end{eqnarray}
in which
\begin{eqnarray*}
\Delta r&=&-\frac{\Pi_{WW}^{(1)}(0)}{M_W^2}
  +\frac{g^2}{16\pi^2}\left\{ 4\left(\Delta-\ln M_Z^2\right)
  +6+\left(4+c_\theta^2-\frac{6c_\theta^2}{s_\theta^2}\right)
     \ln\frac{1}{c_\theta^2}\right\}\\
           &   &\qquad+\frac{g^2}{16\pi^2}\frac{5c_\theta^4-3s_\theta^4}
   {s_\theta^2}\ln\frac{1}{c_\theta^2}-\frac{\delta M_W^2}{M_W^2}
                                            +2\frac{\delta g}{g}.
\label{eq:deltar}
\end{eqnarray*}

To thus accuracy the on-shell $Z^0$ mass is given by
\begin{equation}
(M_Z^{\rm OS})^2=M_Z^2+\Re\Pi_{ZZ}^{(1)}(M_Z^2)+\delta M_Z^2.
\label{eq:lastrc}
\end{equation}

In these expressions $M_W$, $M_Z$, $g$ and $g^\prime$
are the renormalized parameters of the model in what ever renormalization
scheme is being considered. $\Pi_{WW}(q^2)$, $\Pi_{ZZ}(q^2)$,
$\Pi_{\gamma\gamma}(q^2)$ and $\Pi_{Z\gamma}(q^2)$ are the
transverse pieces of the one-particle irreducible $W$, $Z^0$ and
photon self-energies and the $Z$-photon mixing respectively.
The prime denotes differentiation with respect to $q^2$ and
the superscript ${}^{(1)}$ indicates one-loop order.
The sine and cosine of the weak mixing angle are denoted
$s_\theta$ and $c_\theta$, and in all
schemes $\sin^2\theta_W=1-M_W^2/M_Z^2$, where again the masses are
the renormalized masses in the particular scheme of interest.
The quantities
$\delta M_W^2$, $\delta M_Z^2$, $\delta g$ and $\delta g^\prime$ are
mass and coupling constant counterterms which will differ by finite
pieces between renormalization schemes.

Note that $\Delta r$ that appears in eq.(\ref{eq:deltar}) has the same form
for any renormalization scheme. An expression for $\Delta r$
in the $\overline{\rm MS}$ scheme has been given in
ref.\cite{Sirlin2} but that the one of eq.(\ref{eq:deltar})
is both simpler in form and more general. This is because the authors of
ref.\cite{Sirlin2} express the $\overline{\rm MS}$ scheme $\Delta r$
in terms of the on-shell scheme renormalized masses instead of
$\overline{\rm MS}$ scheme masses as is more natural and appropriate.

Each of the equations (\ref{eq:firstrc})--(\ref{eq:lastrc})
is of the same form in that a model-independent physical
observable that must be obtained from experiment appears on the left hand
side and an expression in terms of the unphysical renormalized
parameters appears on the right.

The hadronic uncertainty arises entirely because of the presence of
$\Pi_{\gamma\gamma}^\prime(0)$ in eq.~(\ref{eq:firstrc}). If this equation
can be substituted for by an alternative piece of experimental input
then the hadronic uncertainly will be eliminated.

In the on-shell renormalization one fixes the counterterms by requiring
that the real part of the dressed $W$ and $Z^0$ boson propagators vanish
when evaluated at the renormalized masses $q^2=M_W^2$ and $q^2=M_Z^2$
respectively. The radiative corrections to Thomson scattering are
also required to vanish. At the one-loop this leads to
\begin{eqnarray}
\delta M_W^2&=&-\Re\Pi_{WW}^{(1)}(M_W^2)\label{eq:firstct}\\
\delta M_Z^2&=&-\Re\Pi_{ZZ}^{(1)}(M_Z^2)\\
\frac{\delta g}{g}&=&-\frac{1}{2}\Pi_{\gamma\gamma}^{(1)\prime}(0)
            -\frac{s_\theta}{c_\theta}\frac{\Pi_{Z\gamma}^{(1)}(0)}{M_Z^2}
            -\frac{c_\theta^2}{2s_\theta^2}
                  \Re\left(\frac{\Pi_{WW}^{(1)}(M_W^2)}{M_W^2}
                          -\frac{\Pi_{ZZ}^{(1)}(M_Z^2)}{M_Z^2}\right)\\
\frac{\delta g^\prime}{g^\prime}&
         =&-\frac{1}{2}\Pi_{\gamma\gamma}^{(1)\prime}(0)
           -\frac{s_\theta}{c_\theta}\frac{\Pi_{Z\gamma}^{(1)}(0)}{M_Z^2}
           +\frac{1}{2}\Re\left(\frac{\Pi_{WW}^{(1)}(M_W^2)}{M_W^2}
                          -\frac{\Pi_{ZZ}^{(1)}(M_Z^2)}{M_Z^2}\right).
\label{eq:lastct}
\end{eqnarray}
In the $\overline{\rm MS}$ scheme the counterterms are taken to be just the
divergent pieces of the right hand sides of
equations~(\ref{eq:firstct})--(\ref{eq:lastct}) that introduces
dependence on the arbitrary renormalization scale, $\mu$.

Note that the renormalization  conditions stated in the preceding
paragraph do not represent physical constraints. They are cast in
terms of ``dressed propagators'' and ``radiative corrections''
that are constructs of perturbation theory. They therefore
represent a set of conventions specific to a particular scheme
defining how the perturbation expansion is to be constructed.
Given the renormalization conditions one may solve the
equations~(\ref{eq:firstrc})--(\ref{eq:lastrc})
for the numerical values of the renormalized parameters. The
equations must be solved simultaneously. This approach
was advocated implemented in the program {\tt Z0POLE} \cite{Z0POLE}.
Solving only the first
while ignoring the other two, as is usually done, can  be
inconsistent with constraints from experimental data.

Taking as experimental input the values given in
equations~(\ref{eq:firstexp})--(\ref{eq:lastexp})
and $\mu=M_Z^{\rm OS}$ and solving simultaneously using the program
{\tt Z0POLE} \cite{Z0POLE} yields
\[
\alpha^{-1}_{\overline{\rm MS}}(\mu^2=M_Z^2)=128.17
\]
for $m_t=176$\,GeV, $M_H=100$\,GeV and using the hadronic
contribution given in ref.\cite{Jegerlehner}.
Note that this is not identical to the running $\alpha$ that
is often discussed
\[
\alpha(q^2)=\frac{\alpha(0)}
{1-\left(\Re\Pi_{\gamma\gamma}(q^2)/q^2
                               -\Pi_{\gamma\gamma}^\prime(0)\right)}.
\]
for which $\alpha^{-1}(M_Z^2)=128.90$.
The latter has no firm theoretical basis in the full electroweak
Standard Model and cannot be used in
any sort of self-consistent perturbation expansion.

Once the numerical values of the parameters are fixed, predictions
for all other physical observables may be calculated. In the table
predictions for a number of precisely measured observables along
with their associated error due to the hadronic uncertainty are
given. The
latter has been reevaluated from experimental data again recently
\cite{Jegerlehner,Swartz} and may be is expressed as
\begin{equation}
\Delta\alpha^{(5)}=\Re\frac{\Pi_{\gamma\gamma}^{\rm(h)}(M_Z^2)}{q^2}
-\Pi_{\gamma\gamma}^{{\rm(h)}\prime(0)}=-0.0280\pm0.0007.
\label{eq:hadronunc}
\end{equation}
The superscript
${}^{(h)}$ is meant to indicate the hadronic contribution coming
from Feynman diagrams containing $u$, $d$, $c$, $s$ and $b$ quark loops.
The result quoted in
eq.~(\ref{eq:hadronunc}) is the one given in ref.~\cite{Jegerlehner}.
Central values differ by 0.0015 between groups representing
an unaccounted for component of the hadronic uncertainty.

The tabulated error represents the accuracy that needs to be achieved
for the quantity concerned to become a viable substitute for
$\alpha$ as experimental input. The quantities listed are the $W$ mass,
$M_W$; the $Z^0$ total width, $\Gamma_Z$; the ratio of the hadronic to muon
cross-section, $R$; the left-right polarization asymmetry, $A_{LR}$;
the muon and $b$-quark forward-backward  asymmetries, $A_{FB}^\mu$ and
$A_{FB}^b$; and the $Z^0$ partial width to $b$-quarks, $\Gamma_b$.
All ratios of cross-sections are calculated using the full
observed cross-section with no attempt having been made to split off
non-resonant background contributions.
\bigskip
\begin{center}
\begin{tabular}{|l||c|c|} \hline
        & $\Delta\alpha^{(5)}$ & $\Delta m_t$ \\ \hline\hline
$M_W$ & $\pm13$\,MeV & $\pm30$\,MeV \\ \hline
$\Gamma_Z$ & $\pm0.7$\,MeV & $\pm0.7$\,MeV \\ \hline
$R$ & $\pm0.004$ & $\pm0.003$ \\ \hline
$A_{LR}$ & $\pm0.002$ & $\pm0.0008$ \\ \hline
$A_{FB}^\mu$ &  $\pm0.0004$ & $\pm0.0002$ \\ \hline
$A_{FB}^b$ &  $\pm0.001$ & $\pm0.0006$ \\ \hline
$\Gamma_b$ & $\pm0.1$\,MeV  & $\pm0.7$\,MeV \\ \hline
\end{tabular}
\end{center}
\bigskip

In particular note that if the accuracy of $\pm13$\,MeV could be
attained on the measurement of the $W$ boson mass then it replaces
$\alpha$. If the 0.0015 between the central values of the hadronic
uncertainty persists then a $\pm20$\,MeV measurement of $M_W$ becomes
competitive with $\alpha$.
This is theoretically attractive because of the
simplicity of implementation. The equation~(\ref{eq:firstrc}) would
be replaced by
\begin{equation}
(M_W^{\rm OS})^2=M_W^2+\Pi_{WW}^{(1)}(M_W^2)+\delta M_W^2
\end{equation}
(if one insists on following the same unphysical conventions
used for $M_Z$) and solve the simultaneous equations as before.
Other quantities could be used but their simultaneous solution
is somewhat more complicated to implement. Of these the left-right
polarization asymmetry, $A_{LR}$, looks the most promising.

The table also gives the errors in calculating the listed quantities
by a $\pm5$\,GeV in the top quark mass. The table should therefore
serve a guide to experimentalists at hadron colliders as to the
accuracy required on $M_W$ and $m_t$ in order to make significant
contributions to analyses of electroweak precision data.
There is no significant variation in the given errors with Higgs mass.

{}From the table it may be seen that $M_W$ is measured to $\pm20$\,MeV then
the top quark mass must be measured to $\pm3$\,GeV in order not to pose
an obstacle in precision electroweak analyses.

\section*{Acknowledgements}
This work was supported in part by the U.S.\ Department of Energy.

\end{document}